\documentclass[aps,prep,nofootinbib]{revtex4}
\usepackage{amssymb}
\usepackage{amsmath}
\usepackage{bm}

\begin{document}

\title{Extended General Relativity: large-scale antigravity and short-scale gravity with $\omega=-1$
from five dimensional vacuum }
\author{$^{1,2}$ Jos\'{e} Edgar Madriz Aguilar\thanks{%
E-mail address: madriz@mdp.edu.ar}, and  $^{2,3}$ Mauricio Bellini \thanks{%
E-mail address: mbellini@mdp.edu.ar} }
\affiliation{$^{1}$ Instituto de F\'isica de la Universidad de Guanajuato, C.P. 37150, Le\'on Guanajuato, M\'exico,\\ \\
$^2$ Departamento de F\'isica, Facultad de Ciencias Exactas y Naturales, Universidad Nacional de Mar del Plata, Funes 3350,
C.P. 7600, Mar del Plata, Argentina.\\  \\
$^3$ Consejo Nacional de Investigaciones Cient\'ificas y T\'ecnicas (CONICET), Argentina.\\
E-mail: madriz@mdp.edu.ar; mbellini@mdp.edu.ar}

\begin{abstract}
Considering a five-dimensional (5D) Riemannian spacetime with a
particular stationary Ricci-flat metric, we obtain in the
framework of the induced matter theory an effective 4D static and
spherically symmetric metric which give us ordinary gravitational
solutions on small (planetary and astrophysical) scales, but
repulsive (anti gravitational) forces on very large (cosmological)
scales with $\omega =-1$. Our approach is an unified manner to
describe dark energy, dark matter and ordinary matter. We
illustrate the theory with two examples, the solar system and the
great attractor. From the geometrical point of view, these results
follow from the assumption that exists a confining force that make
possible that test particles move on a given 4D hypersurface.
\end{abstract}

\maketitle

\vskip .5cm


\section{Introduction}

In much the term antigravity has come to present all those
physical phenomena in which the usual gravitational potential is
modified to accommodate repulsive gravitational forces. This is a
fascinating subject which has many implications from the possible
check of antigravity against experiments to the several
theoretical issues that are involved, the 4D principle of
equivalence and energy conservation among others\cite{nieto}. The
idea of antigravity has been subject of different approaches
through the last decades. Scherk\cite{sch} considered this
phenomenon in the framework of supergravity related to fermionic
generators. In essence, antigravity can be treated as one where
the gravitational and other forces between certain objects in a
field theory can mutually cancel. Antigravity has been studied
from five-dimensional Kaluza-Klein (KK) theory, where the extra
dimension is compact\cite{pollard}. The original version of the KK
theory assures, as a postulate, that the fifth dimension is
compact. A few years ago, a non-compactified approach to KK
gravity known as Induced Matter (IM) theory was proposed by Wesson
and collaborators\cite{3}. In this theory all classical physical
quantities, such as matter density and pressure, are susceptible
of a geometrical interpretation. Wesson's proposal also assumes
that the fundamental 5D space in which our usual spacetime is
embedded, should be a solution of the classical 5D vacuum Einstein
equations: $R_{AB}=0$. The mathematical basis of it is the
Campbell-Magaard theorem\cite{campbell}, which ensures an
embedding of 4D general relativity with sources in a 5D theory
whose field equations are apparently empty. That is, the Einstein
equations $G_{\alpha\beta}=(8\pi G/ c^4)\,T_{\alpha\beta}$ are
embedded perfectly in the Ricci-flat equations $R_{AB}=0$. In
simple terms, Wesson uses the fifth dimension to model matter.
More recently, has been suggested that antigravity can be
originated as the repulsion effect of matter and
antimatter\cite{..}. The relationship between antigravity and
antimatter has been studied also in\cite{barkin}. In the framework
of brane world models, has been suggested that antigravity effects
could be important for very large scales when gravitons are
metastable\cite{rubakov}. The possibility of obtaining an infrared
modification to gravity on cosmological scales from extra
dimensions has been considered in\cite{khoury}. Strong antigravity
has been obtained by compactification on a manifold with flat
directions from a higher-dimensional model\cite{tano}. Very
recently was proposed an experiment to measure antigravity with an
antihydrogen beam\cite{exp}.

On the other hand, when applied to cosmic structure on galactic
and larger scales, standard 4D General Relativity and its
Newtonian weak-field limit fail at describing the observed
phenomenology. To reconcile the theory with observations we need
to assume that $\sim\, 85\,\%$ of the mass is seen only through
its observational effect and that $\sim\, 74\,\%$ of the energy
content of the universe is due to either to an arbitrary
cosmological constant or to a not well defined dark energy fluid.
The cosmological constant problem appears to be so serious as the
dark matter problem. The Einstein equations admit the presence of
an arbitrary constant $\Lambda$. The Friedmann solutions with a
positive $\Lambda$ fit very satisfactorily the observational
evidence of an accelerating universe. The problem arises when one
wishes to attach a physical interpretation to $\Lambda$. Since
observations indicate $\Lambda >0$, the dark energy fluid has
negative pressure. Current observations suggest $\omega =-1$ at
all probed epochs\cite{omega}, so models more sophisticated than a
simple $\Lambda$ could seem in principle unnecessary. However, in the context of quantum field
theory, the $\Lambda$ problem translates into an extreme
fine-tuning problem, because $\rho_v(t_P)/ (\sum\Delta\rho_v
) = (1 + 10^{-108})$ is extremely close to $1$, but not exactly
$1$. This problem would disappear if $\Lambda$ were exactly
zero\cite{wein}. An alternative conclusion we can draw from this
failure is that standard 4D General Relativity must be modified on
these cosmic scales, or, in other words that the equation of state
for matter is not $\omega_m=0$, but could be $\omega_m=-1$. In this
letter we explore this idea from a 5D vacuum state using some
ideas of the STM theory.

\section{The Field Equations on 4D Hypersurfaces}

We start by considering a 5D space-time with a Ricci-flat metric
$g_{ab}$ determined by the line element\cite{rom}
\begin{equation}\label{a1}
dS^{2}=\left(\frac{\psi}{\psi_0} \right)^{2}\left[ c^{2}f(r)dt^{2} - \frac{dr^2}{f(r)}-r^{2}(d\theta ^{2}
+sin^{2}\theta d\phi ^{2})\right]-d\psi^{2},
\end{equation}
where
$f(r)=1-(2G\zeta\psi_{0}/(rc^2))[1+c^{2}r^{3}/(2G\zeta\psi_{0}^{3})]$
is a dimensionless function, $\lbrace t,r,\theta,\phi\rbrace$ are
the usual local spacetime spherical coordinates employed in
general relativity and $\psi$ is the space-like extra dimension
that following the approach of the induced matter theory, will be
considered as non-compact. In this line element $\psi$ and $r$
have length units, $\theta$ and $\phi$ are angular coordinates,
$t$ is a time-like coordinate, $c$ denotes the speed of light,
$\psi_0$ is an arbitrary constant with length units and the
constant parameter $\zeta$ has units of $(mass)(length)^{-1}$.

Now let us to assume that the 5D spacetime can be foliated by the family of hypersurfaces
$\lbrace\Sigma _{0} :\psi=\psi _{0}\rbrace$. On every generic hypersurface $\Sigma_{0}$  the
induced metric is given by the 4D line element
\begin{equation}\label{a2}
dS^{2}_{ind}=c^{2}f(r)dt^{2}-\frac{dr^{2}}{f(r)}-r^{2}(d\theta ^{2}+sin^{2}\theta d\phi ^{2}).
\end{equation}
Given the symmetry properties of the 5D spacetime, it seems
natural to assume that the induced matter on $\Sigma_0$ can be
globally described by a 4D energy momentum tensor of a perfect
fluid $T_{\alpha\beta}=(\rho c^2
+P)U_{\alpha}U_{\beta}-Pg_{\alpha\beta}$ where $\rho(t,r)$ and
$P(t,r)$ are respectively the energy density and pressure of the
induced matter. From the relativistic point of view, observers
that are on $\Sigma_0$ move with $U^{\psi}=0$ [see Sect.
(\ref{traj})]. The Einstein field equations on the hypersurface
$\Sigma _{0}$ for the metric in (\ref{a2}), read
\begin{eqnarray}\label{a3}
r\frac{df}{dr}-1+f&=&-\frac{8\pi G}{c^2} r^{2}\rho,\\
\label{a4} r\frac{df}{dr}-1+f&=&\frac{8\pi G}{c^4} r^{2} P.
\end{eqnarray}
The resulting equation of state is
\begin{equation}\label{a7}
P= -\rho c^2 =-\frac{3c^4}{8\pi G}\frac{1}{\psi _{0}^2},
\end{equation}
which technically corresponds to a vacuum equation of state. On
the other hand, regarding that the metric in (\ref{a2}) has
spherical symmetry, we can associate the energy density of induced
matter $\rho$ to a mass density of a sphere of physical mass
$m\equiv \zeta\psi_0$ and radius $r_0$. If we do that, it follows
that $M$ and the radius $r_0$ of such a sphere are related by the
expression $\zeta=r^{3}_{0}/(2G\psi_{0}^3)$. An immediate
consequence of this expression is that in principle given an
specific value of $r_0$, depending of the value of $\psi _0$ we
could induce a large massive object or a mini-massive object. This
way  we can say that is possible in this case to treat the induced
matter as a massive compact object embedded in a 5D vacuum. Some
information that we can obtain by simple inspection of the metric
(\ref{a1}) is that when $G\zeta = \sqrt{3}/ 9$ there is a single
Schwarzschild radius. In this case the Schwarzschild radius is
$r_{Sch} = {\psi_0/\sqrt{3}} \ge r_0$. When it is greater than the
radius of the sphere of parameter $\zeta$, the compact object has
properties very close to those of a black hole on distances $1 \gg
r/\psi_0 > r_{Sch}/\psi_0$, this condition holds when $G\zeta \le
1/ (2 \sqrt{27})\simeq 0.096225$. For $G\zeta > \sqrt{3}/9$ one
obtains that $f(r)<0$ and there is not Schwarzschild radius. When
$G\zeta \le \sqrt{3}/ 9$ there are two Schwarzschild radius, an
interior $r_{S_i}$ and an exterior one $r_{S_e}$, such that by
definition $f(r_{S_i})=f(r_{S_e})=0$. This last case has very
interesting properties and we will focus on the study of that
properties in some scenarios at astrophysical and cosmological
scales. When we assume that the present universe we live in can be
modeled on the 4D hypersurface $\Sigma _{H_0}:\psi
_{0}=cH_{0}^{-1}$, $H_0$ being $H_{0}=73\,\frac{km}{sec}Mpc^{-1}$
the present Hubble constant, we found that the exterior
Schwarszchild radius $r_{S_e}$ becomes of the order of size of the
Hubble radius which is the size of the present observable
universe. On the other hand as the interior Schwarszchild radius
$r_{S_i}$ depends strongly of the value of $G\zeta$ then when
$G\zeta\ll 1$, the interior Schwarszchild radius $r_{S_i}$
approximates to zero. What makes interesting this case is that an
observer located between these two Schwarzschild radius could be
able to see a compact object with a horizon event determined by
$r_{S_i}$ immersed in our observable universe whose size is
determined by the Hubble horizon given by $r_{S_e}$. This
particular case is our interest and in the preceding sections we
will study in detail more properties of it.

\section{Particle trajectories}\label{traj}

In order to study with more detail properties of the metric
(\ref{a2}), we shall describe the geodesic trajectories of
non-massive and massive test particles. All of these are described
by the equation
\begin{equation}\label{geo}
\frac{dU^a}{dS} +\Gamma^a_{\,\,bc}\,U^b U^c=\phi^a,
\end{equation}
where $U^a={dX^a\over dS}$, $\phi^a$ is a external force and the
five-dimensional velocity conditions are fulfilled respectively
\begin{equation}
g_{ab}U^{a}U^{b}=\epsilon,
\end{equation}
where $\epsilon =0,c^2$ for non-massive and massive test particles
that move on the 5D Ricci-flat metric (\ref{a1}).

From the geodesical point of view, the equation (\ref{geo})
implies that\footnote{In this letter $a,b$ run from $0$ to $4$ and
Greek letters rum from $0$ to $3$.}
\begin{eqnarray}
&& \frac{dU^{\alpha}}{dS} +\Gamma^{\alpha}_{\,\,a b}\,U^{a}
U^{b}=\phi^{\alpha},\\
&& \frac{dU^{\psi}}{dS} +\Gamma^{\psi}_{\,\,a b}\,U^{a}
U^{b}=\phi^{\psi},
\end{eqnarray}
where
\begin{eqnarray}
&&  \phi^{\alpha} = 0, \label{10} \\
&& \phi^{\psi}=\frac{\epsilon}{\psi_0}. \label{force} \label{conf}
\end{eqnarray}
In the eq. (\ref{10}) we have supposed the not existence of an
additional fifth force: $\phi^{\alpha}=0$. The only non-zero force
we shall consider is $\phi^{\psi}$, that plays the role of a
confining force.

\subsection{Non-Massive Particles}\label{nonmas}

For non-massive particles on the metric (\ref{a1}) the condition of 5D null-geodesics: $g_{ab}U^{a}U^{b}=0$,
can be written as
\begin{equation}\label{c1}
\frac{p_{t}^{2}}{c^{2}f(r)}\left(\frac{\psi_{0}}{\psi}\right)^{2}-\left(\frac{\psi}{\psi_{0}}\right)^2\frac{(U^r)^2}{f(r)}
-\frac{p_{\phi}^{2}}{r^{2}}\left(\frac{\psi_0}{\psi}\right)^{2}-(U^{\psi})^{2}=0.
\end{equation}
For a photon moving radially and with no motion along the fifth coordinate the equation (\ref{c1}) gives
\begin{equation}\label{in1}
\frac{dr}{dt}=\pm c f(r).
\end{equation}
For a class of observers located at $\Sigma_{H_0} $ the coordinate function $f(r)$ only remains positive when
$r_{S_i}<r<r_{S_e}$,
thus the metric (\ref{a1}) maintains its signature only in the region given by $r_{S_i}<r<r_{S_e}$. Hence the
metric (\ref{a1}) is
valid on such interval. \\
Expressing the equation (\ref{c1}) in terms of the of the angular coordinate $\phi$ and of the variable $u(\phi)$
it yields
\begin{equation}\label{c2}
\left(\frac{du}{d\phi}\right)^{2}+p_{\phi}^{-2}\left(\frac{\psi}{\psi_0}\right)^{2}(U^{\psi})^{2}-\frac{1}{c^2}
p_{t}^{2}p_{\phi}^{-2}+u^{2}-\frac{2G\zeta\psi_{0}}{c^2}u^{3}-1+p_{\phi}^{-2}\left(\frac{\psi}{\psi_0}\right)^{2}
(U^{\psi})^{2}
[\frac{2G\zeta}{c^2}-(\psi_{0}u)^{-3}](\psi_{0}u)=0.
\end{equation}
If the observer moves with velocity $U^{\psi}=0$ on the
hypersurface $\Sigma _0$, the equation (\ref{c2}) reduces to
\begin{equation}\label{c3}
\left(\frac{du}{d\phi}\right)^{2} - \frac{p_{t}^{2}p_{\phi}^{-2}}{c^2}+u^{2}-\left(\frac{2Gm}{c^2}\right)u^{3}-1=0.
\end{equation}
Notice that $\phi^{\psi} =0$ and the hypersurface $\Sigma_0$ is
totally geodesic\cite{sea} when $U^{\psi}=0$. However, this force
is perpendicular to all possible particle trajectories, so that
the system remains conservative on $\Sigma_0$.

Differentiating with respect to $\phi$ we get simply
\begin{equation}\label{c4}
\frac{d^{2}u}{d\phi^2}+u=\left(\frac{3Gm}{c^2}\right)u^{2}.
\end{equation}
This is the orbit equation for non-massive particles for instead
photons. Clearly the expression (\ref{c4}) remains
the same as the deflection equation for light obtained in standard
4D general relativity and therefore we can say that in this formalism the photons deflection
usually obtained in general relativity can be integrally recovered on
$\Sigma _0$.

\subsection{Massive Particles}

For a massive test particle outside of a spherically symmetric compact
object in 5D with exterior metric given by (\ref{a1}) the 5D Lagrangian can be written as
\begin{equation}\label{b1}
^{(5)}L=\frac{1}{2}g_{ab}U^{a}U^{b}=\frac{1}{2}\left(\frac{\psi}{\psi_0}\right)^{2}\left[c^{2}f(r)
\left(U^t\right)^2-\frac{\left(U^r\right)^2}{f(r)}-r^{2}\left(U^{\theta}\right)^2-r^{2}sin^{2}
\theta\left(U^{\phi}\right)^2\right]
- \frac{1}{2}\left(U^{\psi}\right)^2.
\end{equation}
As it is usually done in literature we can, without lost of
generality, confine the test particle to orbits with
$\theta=\pi/2$. Thus, from the Lagrangian $(\ref{b1})$, it can be
easily seen that only $t$ and $\phi$ are cyclic coordinates, so
their associated constants of motion $p_{t}$ and $p_{\phi}$, are
\begin{eqnarray}\label{b2}
p_{t}\equiv\frac{\partial\,^{(5)}L}{\partial U^t}&=&c^{2}\left(\frac{\psi}{\psi_0}\right)^{2}f(r)U^t,\\
\label{b3} p_{\phi}\equiv\frac{\partial\,^{(5)}L}{\partial
U^{\phi}}&=&-\left(\frac{\psi}{\psi_0}\right)^{2}r^{2} U^{\phi}.
\end{eqnarray}
The five-velocity condition gives
\begin{equation}\label{b4}
c^{2}\left(\frac{\psi}{\psi_0}\right)^{2}f(r)\left( U^{t}\right)^{2}-
\left(\frac{\psi}{\psi_0}\right)^{2}\frac{\left(
U^{r}\right)^{2}}{f(r)}-\left(\frac{\psi}{\psi_0}\right)^{2}r^{2}\left(
U^{\phi}\right)^{2} - \left( U^{\psi}\right)^{2} = c^2 .
\end{equation}
By expressing the equation (\ref{b4}) in terms of the constants of motion given by (\ref{b2}) and (\ref{b3}) we obtain
\begin{equation}\label{b5}
\left(\frac{\psi_0}{\psi}\right)^{2}\frac{p_{t}^{2}}{c^{2}f(r)}-\left(\frac{\psi}{\psi_0}\right)^{2}\frac{\left(
U^{r}\right)^{2}}{f(r)}-\frac{p_{\phi}^2}{r^2}\left(\frac{\psi}{\psi_0}\right)^{2}-\left(
U^{\psi}\right)^{2}=c^{2}.
\end{equation}
After rearranging some terms and using the expression for $f(r)$,
the equation (\ref{b5}) can be written as
\begin{equation}\label{b6}
\frac{1}{2}\left(
U^{r}\right)^{2}+\frac{1}{2}\left(\frac{\psi_0}{\psi}\right)^{2}\left(
U^{\psi}\right)^{2} + V_{eff}(r) = E,
\end{equation}
where the effective potential $V_{eff}(r)$ and the total energy $E$ are given
by the expressions
\begin{eqnarray}\label{b7}
V_{eff}(r)&=&-\left(\frac{\psi_0}{\psi}\right)^{2}\frac{G\zeta\psi
_0}{r}+\left(\frac{\psi_0}{\psi}\right)^{4}\left[\frac{p_{\phi}^2}{2r^2}-\frac{G\zeta\psi
_0 p_{\phi}^2}{c^{2}r^3}\right] -
\frac{1}{2}\left(\frac{\psi_0}{\psi}\right)^{2}\left[\left(U^{\psi}\right)^2\left(\frac{2G\zeta\psi_0}{c^2
r}-\frac{r^2}{\psi_0^2}\right)-
\left(\frac{rc}{\psi_0}\right)^{2}\right]\\
\label{b8} E&=&
\frac{1}{2}\left(\frac{\psi_0}{\psi}\right)^{4}(p_{t}^{2}c^{-2}+p_{\phi}^{2}\psi_{0}^{-2})-
\frac{c^2}{2}\left(\frac{\psi_0}{\psi}\right)^{2}.
\end{eqnarray}
When one takes $U^{\psi}=0$, the induced potential $V_{ind}(r)$ on
the hypersurface $\Sigma _0$ is given by
\begin{equation}\label{b9}
V_{ind}(r)=-\frac{G m}{r}+\frac{p_{\phi}^{2}}{2r^2}-\frac{G m}{c^2}\frac{
p_{\phi}^2}{r^3}-\frac{c^2}{2}\left(\frac{r}{\psi_0}\right)^{2},
\end{equation}
where $m=\zeta\psi _0$ is the effective 4D physical mass.The
confining force $\phi^{\psi}=\epsilon/\psi_0$ is given by eq.
(\ref{force}), and the system is conservative on $\Sigma_0$,
because $\phi^{\psi}$ is perpendicular to the penta-velocities
$U^{\mu}$ on all the hypersurface $\Sigma_0$. Hence $\phi^{\psi}$
cannot be interpreted as a fifth force.

The first two terms in the right hand side of (\ref{b9})
correspond to the classical potential, the third term is the usual
relativistic contribution and the last term is a new contribution
coming from the 5D metric solution (\ref{a1}). The Newtonian
acceleration associated to the induced potential (\ref{b9}) reads
\begin{equation}\label{in2}
a=-\frac{Gm}{r^2}+\frac{p_{\phi}^2}{r^3}-\frac{3Gm}{c^2}\frac{p_{\phi}^2}{r^4}+\frac{rc^{2}}{\psi _0^2}.
\end{equation}
In order to describe purely gravitational effects, let us to consider the case when $p_{\phi}=0$.
In this case, the equation (\ref{in2}) reduces to
\begin{equation}\label{dc1}
a=-\frac{Gm}{r^2}+\frac{rc^2}{\psi_{0}^2}.
\end{equation}
This acceleration becomes null when
\begin{equation}\label{dc2}
r_{ga}=\left(\frac{Gm}{c^2}\,\psi_{0}^2\right)^{1/3}.
\end{equation}
By simple inspection of the equation (\ref{dc1}) it can be easily
seen that for $0<r<(Gm\psi_{0}^{2}/c^{2})^{1/3}$ the acceleration
$a$ is negative, which means that the force acting on the test
particle is attractive. For distances
$r>(Gm\psi_{0}^{2}/c^{2})^{1/3}$ the acceleration experimented by
the particle is positive and so it is the force, meaning this that
on this range the test particle is experimenting a repulsive
force. This fact can be interpreted as before
$r=(Gm\psi_{0}^{2}/c^{2})^{1/3}$ the force is gravitational in
nature and after this value the force becomes antigravitational.
We define the radius on which this force is null as the
gravitational-antigravitational radius, and will denote it by
$r_{ga}$. A particular limit, which is very interesting because
one recover the Schwarszchild case, is obtained when we take the
limit $\psi_0 \rightarrow \infty$. In this case
\begin{displaymath}
\lim_{\psi_0 \rightarrow \infty}{ r_{ga}} \rightarrow \infty,
\end{displaymath}
and the effective 4D force on massive particles is purely
attractive. Furthermore, as in standard general relativity one
obtains the equation of state for matter:
$\left.P\right|_{\psi_0\rightarrow\infty} = -\left.\rho
c^2\right|_{\psi_0\rightarrow\infty}=0$.

The condition for circular
motion of the test particle $(dV_{ind}/dr)=0$ acquires the form
\begin{equation}\label{b10}
r^{5}-\frac{Gm}{c^2}\psi_{0}^{2}r^{2}+\frac{p_{\phi}^{2}\psi_{0}^2}{c^2}r-\frac{3Gm}{c^4}p_{\phi}^{2}\psi_{0}^{2}=0.
\end{equation}
By expressing the equation (\ref{b6}) as a function of the angular
coordinate $\phi$ (indeed assuming $r=r(\phi)$), we obtain
\begin{eqnarray}
\left(\frac{du}{d\phi}\right)^{2}&+&(1-\frac{2G \zeta \psi_0}{c^2}
u)(u^{2}+(\psi/\psi _0)^{2}p_{\phi}^{-2})
+\frac{(U^{\psi})^2}{p_{\phi}^2}\left(\frac{\psi}{\psi_0}\right)^{2}+\frac{(U^{\psi})^2}{p_{\phi}^2}
\left(\frac{\psi}{\psi
_{0}}\right)^{2}\frac{2G\zeta\psi_{0}}{c^2}
\left(1-\frac{c^{2}(\psi_{0}u)^{-3}}{2G\zeta}\right)u \nonumber \\
&-& c^{2}(\psi/\psi
_0)^{2}p_{\phi}^{-2}(u\psi_0)^{-2}=c^{-2}p_{t}^{2}p^{-2}_{\phi}+\psi_{0}^{-2},
\label{b11}
\end{eqnarray}
where we have introduced the new variable $u=1/r$. This expression is the equation of motion for a massive
test particle in the metric (\ref{a1}). When we go down from 5D to 4D the expression (\ref{b11}) yields
\begin{equation}\label{b12}
\left(\frac{du}{d\phi}\right)^{2}+(1-\frac{2G m}{c^2}
u)(p_{\phi}^{-2}+u^{2})-c^{2}p_{\phi}^{-2}(u\psi_0)^{-2}=c^{-2}p_{t}^{2}p^{-2}_{\phi}+\psi_{0}^{-2}.
\end{equation}
In order to simplify the algebraic structure of (\ref{b12}) we
derive it with respect $\phi$, obtaining
\begin{equation}\label{b13}
\frac{d^{2}u}{d\phi^2}+u+c^{2}p_{\phi}^{-2}\psi_{0}^{-2}u^{-3}=\left(\frac{G m}{c^2}\right)
p_{\phi}^{-2}+\left(\frac{3G m}{c^2}\right) u^{2}.
\end{equation}
This equation is almost the same that the one usually obtained in
the 4D general theory of relativity for an exterior Schwarzschild
metric with the exception of the third term on the left hand side.
This new term could be interpreted as a new contribution coming in this case
from the extra coordinate.

\section{Some Applications}

To illustrate the approach here presented, we shall study two
examples which are of astrophysical interest. As a first
application we consider a star of one solar mass and in second
place we will consider an object of $6\times 10^{15}$ solar
masses, which is the mass estimated for the core of the great
attractor. In both cases we regard $\psi_{0}=c\,H_{0}^{-1}$ which
means that in this case the size of the fifth dimension
corresponds to the present Hubble radius. For simplicity, we shall
consider the case $p_{\phi}=0$.

\subsection{Solar system}

The core of our solar system is the sun, which has a mass
$m=M_{\odot}$. Its mass is approximately the $98\,\%$ of all the
mass of the solar system. On the other hand, the Oort cloud is a
spherically symmetric cloud of comets which surround the core of
the solar system and has a maximum size of approximately $1\times
10^{5}\,AU\simeq 1.5\times 10^{16}\, mts$\cite{Nihei}. It defines
the outer gravitational limit of the solar system. In order to
test our theory we shall calculate the radius $r_{ga}$, which
should be bigger than this gravitational limit.

The expression (\ref{dc2}) and the metric function $f(r)$ become
\begin{equation}\label{e1}
r_{ga}=\left(\frac{Gm}{c^3}H_{0}\right)^{1/3}\frac{c}{H_{0}}\,,
\end{equation}
\begin{equation}\label{e2}
f(r)=1-\frac{2Gm}{rc^2}-\frac{H_{0}^{2}r^2}{c^2}.
\end{equation}
Employing the equations (\ref{e1}) and (\ref{e2}) and using the
observables $H_{0}=73\,(km/sec)Mpc^{-1}$, $c=299792458\,(mts/sec)$
and $G=6.6743\cdot 10^{-11}\,mts^{3}/(kg.sec^2)$ we obtain that
for a star of mass $m=1M_{\odot}$ the gravitational and
antigravitational radius is given approximately by
$r_{ga}=2.8771\cdot 10^{18}\,mts$ which in parsecs is
$r_{ga}=93.24\, pc$. The interior Schwarzschild radius for this
case is $r_{s_{i}}=2953.2359\,mts$ which coincides exactly with
the usual Schwarzschild radius obtained in 4D general relativity.
The exterior Schwarzschild radius is $r_{s_{e}}=1.27\cdot
10^{26}\,mts$ which coincides with the size of the observable
universe i.e. it coincides with the present Hubble horizon. With
these data the picture we can get from our solar system is that
for length scales greater than $93.24\, pc$ the gravity of the sun
will present a different face in this case antigravity will be
present. This result is compatible with observable structures
trapped by the gravitational field of the sun (the Oort cloud
which is located approximately to $1 \times 10^{5}\, AU$), which
are approximately two orders of magnitude below $r_{ga}$. Notice that this is only an estimation, because we are
considering $p_{\phi}=0$.\\

\subsection{Great attractor}

The Great Attractor is the largest and most important cluster
concentration of galaxies in the local Universe. Although the
motion of galaxies towards the GA is significant, there is no
evidence for any backside infall onto the GA. This suggests that
the galaxy flow in this region is just a part of a larger flow,
caused by some more massive attractive center. Its mass can be
estimated in $m=6\times 10^{15}\,M_{\odot}$\cite{bohe}, the radius
$r_{ga}$ take the value $r_{ga}=5.22808\times 10^{23}\,mts\,
\simeq 16.94305\,Mpc$. The exterior Schwarzschild radius is
$r_{s_e}=1.26699\cdot 10^{26}\,mts$ which is the present day
Hubble horizon. Our results for the gravitational influence of the
core of the SC are in very good agreement with observation. The
$r_{ga}$ estimated in our model for the great attractor is of the
same order of magnitude than the observed one. For distances
grater than this radius antigravitational effects will appear
according to our model. These antigravitational effects  could
explain why at until astrophysical scales 4D general relativity
works predicting purely gravitational interactions and why at
cosmological scales the effects of an accelerated expansion can be
affecting matter distributions.

\section{Conclusions}

In this letter we have studied on the framework of an extended version of General Relativity a 5D vacuum state solution given by the Ricci-flat metric (\ref{a1}). In our approach, matter is considered as a $\omega =
p_m/\rho_m= -1$ 4D vacuum state, such that the pressure on the
effective 4D manifold is $P=-3 c^4/( 8\pi G \psi^2_0)$, being
$\psi_0=c/H_0$ the Hubble radius. The effective 4D metric
(\ref{a2}), is static, exterior and describes spherically
symmetric matter (ordinary matter, dark matter and dark energy) on
scales $r_0 < r_{Sch} < c/H_0$ for black holes or $r_{Sch} < r <
c/H_0$ for ordinary stars with $r_0$ being the radius of the star. The
metric (\ref{a2}) describes both, gravity (for $r < r_{ga}$) and
antigravity (for $r > r_{ga}$), $r_{ga}$ being the radius for
which the effective Newtonian acceleration (\ref{dc1}) becomes
zero. Notice that in that calculation we have considered null angular
momentums $p_{\phi}=0$, fact that allow us to interpret the repulsive force acting on a
test particle as a genuine antigravitatory effect. The radius $r_{ga}$ is very
important because delimitates distances for which dark energy and
ordinary matter (dark matter and ordinary matter) are dominant: $r>
r_{ga}$ ($r < r_{ga}$). In resume we conclude that all of these,
ordinary matter, dark matter and dark energy can be considered as
matter subject to a generalized gravitational field which is
attractive on scales $r< r_{ga}$ and repulsive on scales $r >
r_{ga}$. In simple words gravity from a 5D vacuum state can have two facets, an attractive
nature known as gravity and a repulsive one that we interpret as antigravitational, every one
acting on different length scales: astrophysical and cosmological respectively.

\section*{Acknowledgements}

\noindent J.E.M.A acknowledges CONACYT M\'exico and M.B.
acknowledges UNMdP and CONICET Argentina for financial support.

\end{document}